%% file: main.tex
\documentclass[journal]{IEEEtran}
\usepackage{hyperref}
\hypersetup{colorlinks = true, allcolors = black}
\usepackage{enumitem}
\usepackage{amsmath,amsfonts}

\usepackage{algorithmic}
\usepackage{graphicx}
\usepackage{textcomp}
\usepackage{xcolor}

\usepackage{tikz}
\usepackage{enumitem}
\setlist{nosep}
\usepackage[ruled]{algorithm2e}
\usepackage{booktabs}
\usepackage{url}
\usepackage{multirow}
\usepackage{verbatimbox}
\usepackage[labelfont=bf]{caption}

\ifCLASSINFOpdf

\else

\fi

\begin{document}
%
\title{\textsc{threaTrace}: Detecting and Tracing Host-based Threats in Node Level Through Provenance Graph Learning}
%
%
%

\author{Su~Wang,
        Zhiliang~Wang,~\IEEEmembership{Member,~IEEE,}
        Tao~Zhou,
        Xia~Yin,~\IEEEmembership{Senior Member,~IEEE,}
        Dongqi~Han,
        Han~Zhang,
        Hongbin~Sun,
        Xingang~Shi,~\IEEEmembership{Member,~IEEE,}
        Jiahai~Yang,~\IEEEmembership{Senior Member,~IEEE}
}


%
%

\markboth{Journal of \LaTeX\ Class Files,~Vol.~14, No.~8, August~2015}%
{Shell \MakeLowercase{\textit{et al.}}: Bare Demo of IEEEtran.cls for IEEE Journals}
%



\maketitle

\begin{abstract}
Host-based threats such as Program Attack, Malware Implantation, and Advanced Persistent Threats (APT), are commonly adopted by modern attackers. Recent studies propose leveraging the rich contextual information in data provenance to detect threats in a host. Data provenance is a directed acyclic graph constructed from system audit data. Nodes in a provenance graph represent system entities (e.g., $processes$ and $files$) and edges represent system calls in the direction of information flow. However, previous studies, which extract features of the whole provenance graph, are not sensitive to the small number of threat-related entities and thus result in low performance when hunting stealthy threats.

We present \textsc{threaTrace}, an anomaly-based detector that detects host-based threats at system entity level without prior knowledge of attack patterns. We tailor GraphSAGE, an inductive graph neural network, to learn every benign entity's role in a provenance graph. \textsc{threaTrace} is a real-time system, which is scalable of monitoring a long-term running host and capable of detecting host-based intrusion in their early phase. We evaluate \textsc{threaTrace} on three public datasets. The results show that \textsc{threaTrace} outperforms three state-of-the-art host intrusion detection systems.
\end{abstract}


%
\IEEEpeerreviewmaketitle

\input{introduction}

\input{relatedwork}

\input{background}

\input{threatmodel}

\input{design}

\input{evaluation}

\input{discussion}

\input{conclusion}



\ifCLASSOPTIONcaptionsoff
  \newpage
\fi



%
\bibliographystyle{IEEEtran}
\bibliography{myReference}

\end{document}

%% file: introduction.tex
\section{Introduction}
\label{section:1}

\IEEEPARstart{N}{owadays}, attackers tend to perform intrusion activities in important hosts of those big enterprises and governments \cite{1}. They usually exploit zero-day vulnerabilities to launch an intrusion campaign in a target host stealthily and persistently, which makes it hard to be detected.

Recent studies \cite{7,8,16,17,18,19, 56, 58} propose leveraging the rich contextual information on data provenance to perform host-based intrusion detection. Compared to the raw system audit data, data provenance contains richer contextual information, which is useful for separating the long-term behavior of threats and benign activities \cite{19,20}.

Some anomaly-based methods \cite{18,19} use graph kernel algorithm to dynamically model the whole graph and detect abnormal graphs by clustering approaches. However, the provenance graph of a system under stealthy intrusion campaigns may be similar to those of benign systems. Therefore, leveraging graph kernel to extract the features of whole graphs is not sensitive to a small number of anomalous nodes in the graph. The graph-kernel-based methods also lack the capability to locate the position of anomalous nodes, which is essential to trace abnormal behavior and fix the system. \cite{56} focuses on hunting malware by detecting anomalous paths in a provenance graph. However, some complex threats (e.g., APT) split their campaign into several parts instead of appearing in a complete path, which makes it difficult to be detected by path-level methods. Misuse-based methods \cite{7,8, 58} typically define some attack patterns and use them to match anomalous behavior. In consideration of the frequent use of zero-day exploits by modern attackers, misuse-based methods lack the ability to detect unknown threats.

We present \textsc{threaTrace}, an anomaly-based detector that is capable of efficiently detecting stealthy and persistent host-based threats at node level. \textsc{threaTrace} takes data provenance as source input and tailors an inductive graph neural network framework (GraphSAGE) \cite{23} to learn the rich contextual information in data provenance. GraphSAGE is a kind of Graph Neural Network (GNN), which is a group of neural network models designed for various graph-related work and has succeeded in many domains (e.g., Computer Vision \cite{25,26}, Natural Language Processing \cite{27,28}, Chemistry and Biology \cite{29,30}, etc.). We utilize its ability to learn structural information about a node's role in a provenance graph \cite{23} for node-level threats detection.

Using GraphSAGE for threats detection is challenging: 1) How to train the model without prior knowledge of attack patterns? 2) How to tackle the data imbalanced problem \cite{52}?
	
We tailor a novel GraphSAGE-based framework that tackles those challenges. Different from the previous graph level detection methods, \textsc{threaTrace} learns every benign node's role in a system data provenance graph to capture stealthy abnormal behavior without prior knowledge of attack patterns. We design a multi-model framework to learn different kinds of benign nodes, which tackles the problem of data imbalance and effectively improves the detection performance. \textsc{threaTrace} is a real-time system that can be deployed in a long-term running system with acceptable computation and memory overhead. It is capable of detecting host intrusions in their early phase and finding the location of anomalous behavior. We evaluate \textsc{threaTrace} in three public datasets and demonstrate that \textsc{threaTrace} can effectively detect host-based threats with a small proportion in the whole-system provenance with fast processing speed and acceptable resource overhead. 

Our paper makes contributions summarized as follows:

\begin{itemize}[leftmargin=*]

\item \textbf{Novel node level threats detection}. To the best of our knowledge, \textsc{threaTrace} is the first work to formalize the host-based threats detection problem as an anomalous nodes detection and tracing problem in a provenance graph. Based on the problem statement, we propose a novel GraphSAGE-based multi-model framework to detect stealthy threats at node level.

\item \textbf{High detection performance and novel capability}. We evaluate \textsc{threaTrace}'s detection performance in three public datasets and compare it with three state-of-the-art APTs detection approaches. \textsc{threaTrace} outperforms them in these datasets. We further evaluate \textsc{threaTrace}'s threats tracing ability in a host. Results show that \textsc{threaTrace} can successfully detect and trace the anomalous elements.

\item \textbf{Complete system and open source}. We implement an open-source host-based threats detection system \footnote{\url{https://github.com/threaTrace-detector/threaTrace/}.}.

\end{itemize}

This paper is organized as follows. Related work is introduced in \S\ref{section:8}. Background and motivation of our work are introduced in \S\ref{section:2}. We introduce the threat model in \S\ref{section:3}. An overall introduction of \textsc{threaTrace} is presented in \S\ref{section:4}. We introduce the experiments in \S\ref{section:6} and discuss some issues and limitations in \S\ref{section:7}. We conclude this paper in \S\ref{section:9}.

%% file: relatedwork.tex
\section{Related Work}

\label{section:8}

We study the problems of provenance-based threats detection and anomaly tracing. Thus, we discuss related work in these areas.

\textbf{Provenance-based threats detection.} Data provenance is attractive in the host threats detection community recently, which can be classified as misuse-based and anomaly-based. 

Misuse-based methods detect abnormal behavior based on learning patterns of known attacks. Holmes \cite{8} focuses on the alert generation, correlation, and scenario reconstruction of host-based threats. It matches a prior definition of exploits in a provenance graph based on expert knowledge of existing TTPs (Tactics, Techniques, and Procedures). Poirot \cite{7} detect threats based on correlating a collection of indicators found by other systems and constructs the attack graphs relying on expert knowledge of existing cyber threat reports. It detects threats according to the graph matching of the provenance graphs and attack graphs. It is hard for them to detect unknown threats not included in the TTPs and cyber threat reports. 

Anomaly-based approaches learn models of benign behavior and detect anomalies based on the deviations from the models. StreamSpot \cite{18} proposes to detect intrusion by analyzing information flow graphs. It abstracts features of the graph locally to learn benign models and uses a cluster approach to detect abnormal graphs. Unicorn \cite{19} further proposes a WL-kernel-based method to extract the features of the whole graph, learning evolving models to detect abnormal graphs. Graph kernel methods are used to measure the similarity of different graphs. Unicorn has better performance than StreamSpot using its contextualized graph analysis and evolving models. However, due to the restriction of graph kernel methods, they have difficulty detecting stealthy threats. IPG \cite{75} embeds the provenance graph of each host and reports suspicious hosts based on an autoencoder method. As a graph level approach, IPG has the same limitations as StreamSpot and Unicorn. Log2vec \cite{52} proposes to detect host threats based on anomaly logs detection. It uses logs to construct a heterogeneous graph, extracts log vectors based on graph embedding, and detects malicious logs based on clustering. It is different from provenance-based methods because nodes in a provenance graph are entities of a system instead of logs. ProvDetector \cite{56} is proposed to detect malware by exploring the provenance graph. ProvDetector embeds paths in a provenance graph and uses Local Outlier Factor method \cite{57} to detect malware. Besides malware, host-based threats' behavior is more diverse. Therefore, it is not enough to mine threats from the paths of the provenance graph. Pagoda \cite{60} focuses on detecting host-based intrusion in consideration of both detection accuracy and detection time. In Pagoda's framework, a rule database is required, which is different from \textsc{threaTrace}.

Besides provenance, there are other data sources used for host-based anomaly detection. \cite{55} detects stealthy program attacks by modeling normal program traces and using clustering methods to make anomaly detection. \cite{14} models each state as the invocation of a system call from a particular call site by a finite-state automaton (FSA). However, neither program traces nor system calls are suitable for stealthy threats detections because of the lack of contextual information. \cite{59} detects the lateral movement of APTs taking a graph of authenticating entities as input. Nodes in the authentication graph represent either machines or users. However, this approach cannot detect intrusion campaigns within a single host. 

\textbf{Anomaly tracing.} When deploying an intrusion detection system in a host under monitoring, it is important to trace abnormal behavior instead of only raising alarms. \textsc{threaTrace} uses the node classification framework to directly trace anomalies when detecting them. Anomaly-based state-of-the-art detectors \cite{18,19} raise alarms when the provenance graph of the system is detected as abnormal. However, they cannot trace the position of anomalies. Detectors that are capable of tracing the location of anomalies are usually misuse-based such as Holmes \cite{8} and Poirot \cite{7}, which need prior knowledge about abnormal graph patterns. Nodoze \cite{20} is another approach to identity anomalous path in provenance graphs. Unlike those detectors mentioned above, Nodoze is a secondary triage tool that needs alerts from other detectors as input. Alerts from the current anomaly-based threats detector are not suitable for Nodoze because they are graph-level without detailed information. RapSheet \cite{58} is another secondary triage system that uses alerts from other TTPs-based EDR (Endpoint Detection and Response) and filters false positives in constructed TPGs (Tactical Provenance Graphs). RapSheet faces the same problem as those misuse-based detection methods. SLEUTH \cite{62} performs tag and policy-based attack detection and tag-based root-cause and impact analysis to construct a scenario graph. PrioTracker \cite{63} quantifies the rareness of each event to distinguish abnormal operations from normal system events and proposes a forward tracking technique for timely attack causality analysis. MORSE \cite{61} researches the dependence explosion problem in the retracing of attacker's steps. The attack detection part of MORSE's framework is the same as the SLEUTH \cite{62} system. The attack detection part of SLEUTH \cite{62} and MORSE \cite{61} are both misuse-based. ATLAS \cite{65} proposes a sequence-based learning approach for detecting threats and constructing the attack story. Different from the misuse-based methods \cite{7,8,61,62}, ATLAS uses attack training data to learn the co-occurrence of attack steps through temporal-ordered sequences. The methods introduced above propose various methods to trace an intrusion and construct the attack story. \textsc{threaTrace} can trace the anomaly without a knowledge base or attack training data as input. However, it cannot construct the attack story. We plan to research the gap between anomaly-based methods and attack story construction in future work.

%% file: background.tex
\section{Background \& Motivation}
\label{section:2}

\subsection {Data provenance}

\label{section:2.1}

In recent years, data provenance is proposed to be a better data source for host-based threat detection. It is a directed acyclic graph constructed from system audit data representing the relationship between subjects (e.g., $processes$) and objects (e.g., $files$) in a system. Data provenance contains rich contextual information for threats detection.

\subsection {GraphSAGE}

GraphSAGE \cite{23} is a general inductive GNN framework for efficiently generating node embeddings with node feature information. Unlike the transductive methods, which embed nodes from a single fixed graph, the inductive GraphSAGE learns an embedding function and operates on evolving graphs. Timeliness is an essential factor for preventing intrusion. Therefore, as an inductive GNN approach, GraphSAGE is suitable for performing threats detection via analyzing streaming provenance graphs. GraphSAGE has been proved to be capable of learning structural information about a node's role in a graph \cite{23}. Due to space constraints, we refer the readers with the interest of the theoretical analysis to the paper \cite{23}.

\label{section:2.2}

\begin{figure*}[htbp]
\setlength{\abovecaptionskip}{0.cm}
\setlength{\belowcaptionskip}{-0.3cm}
\centering
\includegraphics[width=1.0\textwidth]{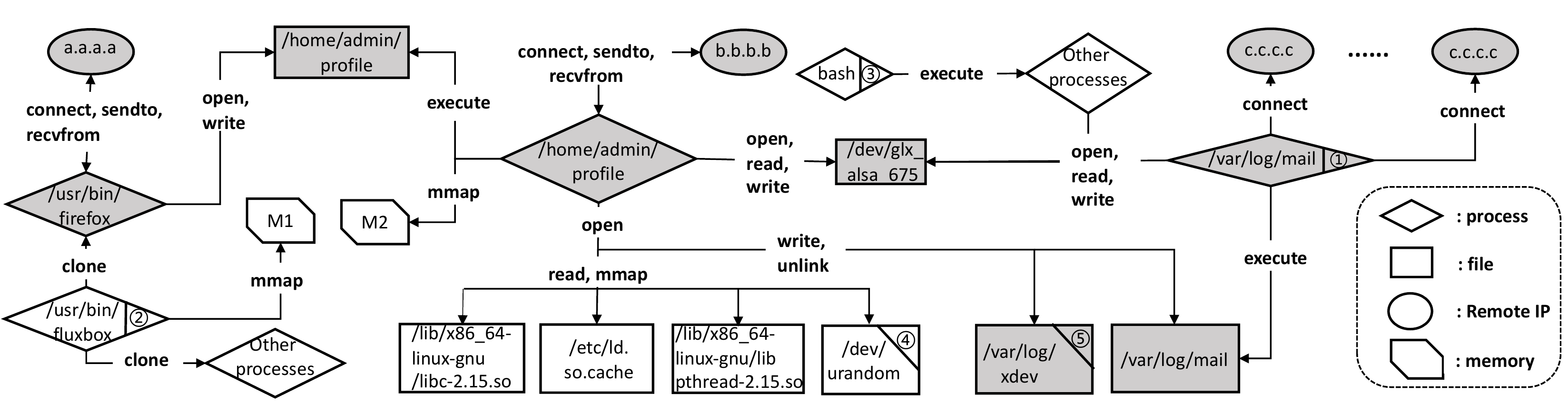}
\caption{A toy example of a provenance graph, which contains host-based intrusion behaviour.}
\label{fig:1}
\vspace{-3mm}
\end{figure*}


\subsection {Motivation Example}

\label{section:2.3}

We present an example (Figure \ref{fig:1}) to illustrate the limitation of state-of-the-art threat detection work and the intuition of our approach. The example is generated from the DARPA TC THEIA experiment \cite{47}. The dark nodes are those nodes related to the intrusion campaign. The attack on the graph lasts for two days. The attacker exploits \texttt{Firefox 54.0.1} backdoor to implant the file named \texttt{/home/admin/profile} in the victim host. It then runs as a process with root privilege to connect with the attacker operator console \texttt{b.b.b.b:80}. A file named \texttt{/var/log/mail} is implanted and elevated as a new process with root privileges. Finally, the attacker carries out a port scan of \texttt{c.c.c.c}. 

\noindent \textbf{Limitations of state-of-the-art threat detection work.} It remains some challenges to detect the threat presented in the example, which leads to limitations of the state-of-the-art threat detection work.

\begin{itemize}[leftmargin=7mm]
	
\hypertarget{P1}{\item[ \textbf{P1} :]} \emph{Rules.} there are some misuse-based \cite{7,8,58} methods for host-based intrusion detection. Rules are important for misuse-based methods, and TTP rules based on MITRE Att\&CK are commonly used. However, it is hard to select a ruleset. Because many MITRE ATT\&CK behaviors are only sometimes malicious \cite{58}, if the rules are macroscopic, it will generate many false alarms. On the contrary, micro rules are hard to detect zero-day attacks.

\hypertarget{P2}{\item[ \textbf{P2} :]} \emph{Stealthy attacker.} Anomaly-based methods Unicorn \cite{19} and StreamSpot \cite{18} use graph kernel algorithm to dynamically model the whole graph and detect abnormal graphs by clustering approaches. However, the intrusion may be stealthy, which means that a system's provenance graph under intrusion campaign may be similar to those of benign systems. In this example, there are millions of benign nodes and less than 30000 anomalous nodes. The proportion of anomalous nodes is less than 1\%, which thus results in a high similarity of the attack graph and benign graph. Therefore, graph-kernel-based methods \cite{18,19} are insensitive to the small anomalous nodes. ProvDetector \cite{56} proposes to select several rarest paths in a graph and detect anomalous paths through their embedding. ProvDetector achieves outstanding performance in malware detection. However, there are more complicated intrusions besides malware. In this example, there are thousands of edges generated by node \textcircled{1}, which brings challenges for path-level detection methods.

\hypertarget{P3}{\item[ \textbf{P3} :]} \emph{Anomaly tracing.} Anomaly-based \cite{18,19} methods modeling the whole graph lack the capability of tracing the location of anomalous behavior, which is vital to trace abnormal behavior and fix the system. In this example, these methods can only raise alarms for the whole graph instead of the specific attack entities (e.g., \texttt{/home/admin/profile}).

\hypertarget{P4}{\item[ \textbf{P4} :]} \emph{Expanding provenance.} Methods \cite{8,18} that store the provenance graph in memory lack the scalability on long-term running systems, which thus results in low practicality.

\end{itemize}

\noindent \textbf{Intuition of our approach.}

The central insight behind our approach is that even for a stealthy intrusion campaign, which tries to hide its behavior, the nodes corresponding to its malicious activities still have different behavior from benign nodes. In Figure \ref{fig:1}, the anomalous process node \textcircled{1} \texttt{/var/log/mail} has thousands of \texttt{connect} edges to remote IP nodes, which is different from a benign process node \textcircled{2} \texttt{/usr/bin/fluxbox}. this phenomenon inspires our work to formalize the host intrusion detection problem as an anomalous nodes detection problem. We tailor a GraphSAGE-based framework to complete a nodes detection task. The detection results of this attack Figure \ref{fig:1} is presented as a case study in \S\ref{section:6.3}. The motivation example is also used to illustrate the design of our approach. We will go back to the example several times in \S\ref{section:4}.

%% file: threatmodel.tex
\section{Threat Model}

\label{section:3}

In this paper, we focus on detecting and tracing anomalous entities in a host caused by intrusion campaigns. We assume the adversary has the following characteristics:

\begin{itemize}[leftmargin=*]

\item \textbf{Stealthy.} Instead of simply performing an attack, the attacker consciously hides their malicious activities, trying to mix their behavior with lots of benign background data, which makes the victim system like a benign mode. 

\item \textbf{Persistent.} The attack tends to last for a long time.

\item \textbf{Frequent use of zero-day exploits.} The attacker tends to use zero-day exploits to attack a system. Therefore, we assume that we do not have any attack patterns for training.

\item \textbf{Has attack patterns in the provenance graph.} In order to complete a malicious activity that is different from the benign activity, the attacker's behavior should leave some attack patterns in the provenance. The attack patterns would make the local structure of an attacker's node different from those of benign nodes with the same label. For example, in Figure \ref{fig:1}, the local structure of the attacker's $process$ node \textcircled{1} is significantly different from the benign $process$ node \textcircled{2}. As another example, a $process$, which never interacts with $files$ before and suddenly starts reading and writing on a $file$, may be an anomalous $process$ that the attacker has controlled. Note that restricted by the granularity of provenance graph, some threats are out of \textsc{threaTrace}'s detection scope. We discussion this limitation in \S\ref{section:7}.

\end{itemize}

We define an \textit{entity} of a running system as benign when there are no malicious activities related to it. An \textit{entity} is defined as abnormal when its behavior is different from a benign entity, which is usually caused by some intrusion campaigns. The purpose of \textsc{threaTrace} is to detect the anomalous entities by monitoring the host and analyzing the nodes on the provenance graph.

We also assume the correctness of the following contents: 1) the provenance collection system; 2) the ability of GraphSAGE to learn structural information about a node's role in a graph, which has been proved in \cite{23}.

%% file: design.tex
\section{Design}

\label{section:4}

\begin{figure*}[htbp]
\setlength{\abovecaptionskip}{0cm}
\centerline{\includegraphics[width=1.0\textwidth]{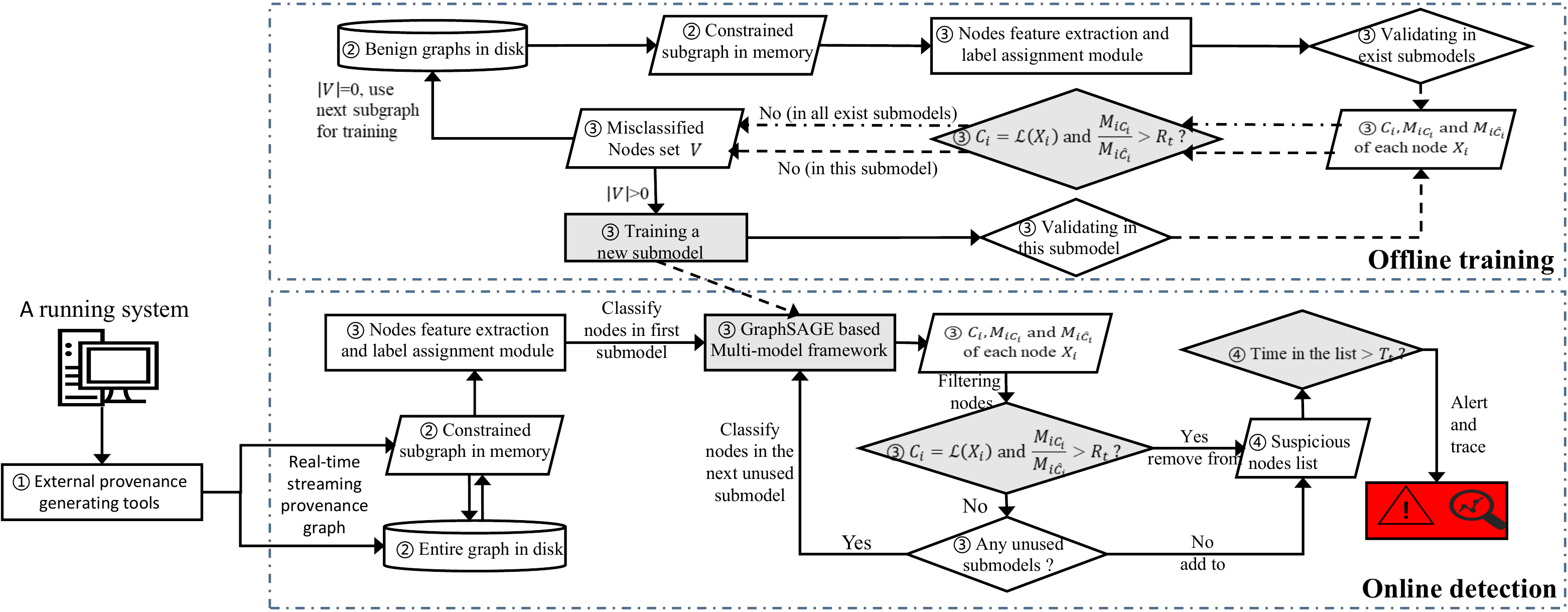}}
\caption{The overview of components in \textsc{threaTrace}. \textcircled{1}: Data Provenance Generator; \textcircled{2}: Data Storage; \textcircled{3}: Model; \textcircled{4}: Alert and Trace.}
\label{fig:2}
\vspace{-3mm}
\end{figure*}

In this section, we present the design of \textsc{threaTrace}. The main components are shown in Figure \ref{fig:2} and presented as follows. We detail them further in the denoted subsections.

\begin{itemize}[leftmargin=*]

\item \textbf{(\S\ref{section:4.1}) Data Provenance Generator.} This component collects audit data of a system in a streaming mode and transforms them into a data provenance graph for the following analysis.

\item \textbf{(\S\ref{section:4.2}) Data Storage.} This component allocates data to the disk and memory. We store the whole graph in the disk to keep the history information and maintain a subgraph with limited size in memory for training and detecting. This storage strategy guarantees the scalability (\hyperlink{P4}{\textbf{P4}}) and dynamic detection capability of \textsc{threaTrace}.

\item \textbf{(\S\ref{section:4.3}) Model.} This is the core component of \textsc{threaTrace}. We use graph data originated and allocated by the previous two components as input and output anomalous nodes.

It is challenging to make ideal abnormal nodes detection: 

\begin{itemize}[leftmargin=7mm, topsep = 6 pt]

\hypertarget{C1}{\item[ \textbf{C1}:]} We assume that we do not have prior attack knowledge in the training phase. Thus we cannot train the model in traditional binary classification mode.

\hypertarget{C2}{\item[ \textbf{C2}:]} Host-based threat detection has a data imbalanced problem \cite{52}. Anomalous nodes may have a small proportion in the provenance graph during execution. Thus, it is more likely to raise false positives that may disturb the judgment and cause ``threat fatigue problem'' \cite{37}.

\hypertarget{C3}{\item[ \textbf{C3}:]} The difference between an anomalous node and benign nodes may not be big enough (e.g., nodes \textcircled{4} an \textcircled{5} in Figure \ref{fig:1}), which thus results in a false negative.

\end{itemize}

In order to tackle these challenges, we tailor a GraphSAGE-based multi-model framework to learn the different classes of benign nodes in a provenance graph without abnormal data (\hyperlink{P1}{\textbf{P1}}, \hyperlink{C1}{\textbf{C1}}). Then we detect abnormal nodes based on the deviation from the predicted node type and its actual type. In this way, \textsc{threaTrace} detects abnormal nodes that have a small proportion in a provenance graph under a stealthy intrusion campaign (\hyperlink{P2}{\textbf{P2}}) and directly locates them (\hyperlink{P3}{\textbf{P3}}). We propose a probability-based method, which reduces false positives and false negatives (\hyperlink{C2}{\textbf{C2}}, \hyperlink{C3}{\textbf{C3}}).

\item \textbf{(\S\ref{section:4.4}) Alert and Trace.} We obtain the abnormal nodes from the previous component and determine whether to raise alerts and trace anomalies in this component. We set a waiting time threshold and a tolerant threshold to reduce false positives and determine whether to raise alerts for the monitored system. Because \textsc{threaTrace} detects threats at node level, we directly trace the position of abnormal behavior in the local neighbor of the abnormal nodes.

\end{itemize}

\subsection {Data Provenance Generator}

\label{section:4.1}

Like many other provenance-based threats detection approaches \cite{7, 8, 18, 19}, we use an external tool Camflow \cite{38} to construct a single, whole-system provenance graph with time order. Camflow provides strong security and completeness guarantees to information flow capture.

\begin{figure}[htbp]
\setlength{\abovecaptionskip}{0.cm}
\setlength{\belowcaptionskip}{-0.3cm}
\centerline{\includegraphics[width=0.3\textwidth]{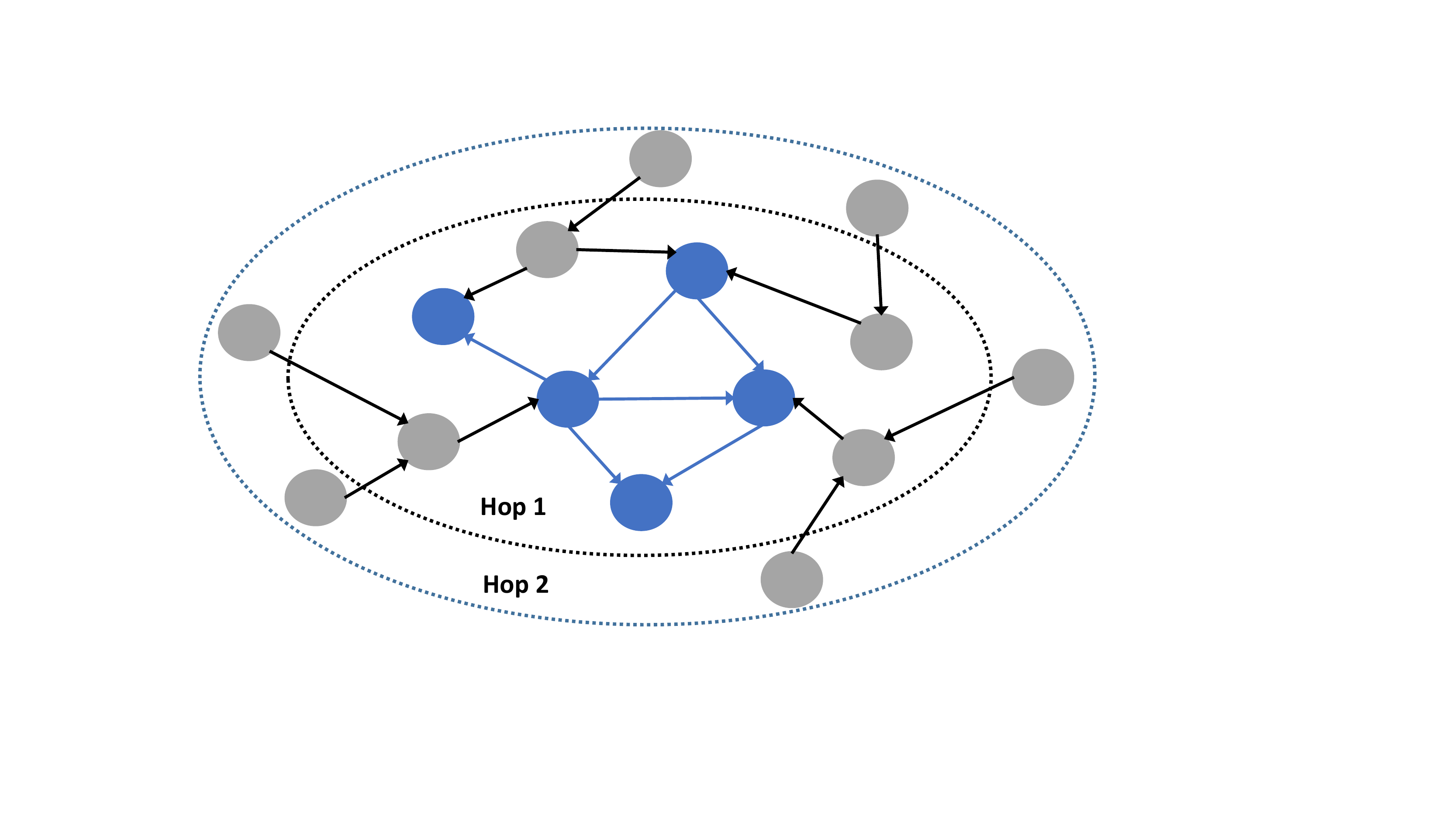}}

\caption{An example of the subgraph maintaining in memory. Nodes in blue denote the $active$ nodes, and the gray nodes denote the $related$ nodes.}
\label{fig:3}
\vspace{-3mm}
\end{figure}

\subsection {Data Storage}

\label{section:4.2}

This module is designed to allocate graph data from the previous module to disk and memory for later usage. In a streaming mode, we append incoming nodes and edges to the whole graph which is stored in the disk. We also maintain a subgraph in memory consisting of $active$ nodes, $related$ nodes, and edges between them. We define $active$ nodes as the entities in the graph that are used to be trained or detected. Nodes that can reach an $active$ node in 2 hops are named as $related$ nodes, which contain history information for training or detecting. Figure \ref{fig:3} presents an example of the subgraph in memory. The subgraph maintaining strategies are different in training and executing phase, which are presented in \S\ref{section:4.3}.

\subsection {Model}

\label{section:4.3}

This is the core component of \textsc{threaTrace}. We design the framework and methods based on the characteristics of host-based threats. 

\textbf{Feature Extraction.} It is challenging to achieve the goal of detecting abnormal nodes without learning attack patterns (\hyperlink{C1}{\textbf{C1}}). The traditional anomaly detection task usually uses benign and abnormal data to train a binary classification model in supervised mode. However, because we assume we do not have prior knowledge about attack patterns during the training phase, we cannot train our model with binary labeled data. GraphSAGE has an unsupervised node classification mode, but it is unfit for threats detection because it assumes that nodes close together have a similar class. In a provenance graph, the close nodes may have different classes (e.g., two close nodes may be a \textit{process} and a \textit{file}). 

Since GraphSAGE's unsupervised mode is unsuitable for threats detection, we tailor a feature extraction and label assignment approach, which allows \textsc{threaTrace} to train the model in a supervised mode without abnormal data and learn every benign node's local structure information. We set a node's label as its node type and extract its features as the distribution of numbers of different edges' types related to it. The GraphSAGE model is trained with labeled data in a supervised mode, which learns different roles of benign nodes. The feature extraction approach is based on the central insight behind our model that the nodes related to the malicious activities have different behavior from the benign nodes that our model learns during the training phase. Thus, if a node is misclassified during the executing phase, it means that its role is different from what it should be. That is, it may be an abnormal node with a malicious task different from the benign nodes. 

The specific extraction process is as follows. We first count the number of different node types (e.g., $file$, $process$, $socket$) and edge types (e.g., $read$, $write$, $fork$) as $N_n$ and $N_e$ before training. We further set a node type and edge type mapping function as $\mathcal{M}_v : \sum \rightarrow \mathbb{N}$ and $\mathcal{M}_e : \sum \rightarrow \mathbb{N}$ to map node type and edge type to an integer range from $0$ to $N_n-1$ and $0$ to $N_e-1$. We learn a function $\mathcal{L} : V \rightarrow \{0, ..., N_n-1\}$ to assign label for $\forall v \in V$ as $\mathcal{L}(v) = \mathcal{M}_v( \mathcal{X}_v(v))$. We learn a function $\mathcal{F} : V \rightarrow \mathbb{N}^{2*N_e}$ to assign features for $\forall v \in V$ as $\mathcal{F}(v) = [a_0, a_1, ..., a_{N_e-1}, a_{N_e}, a_{N_e+1}, ..., a_{N_e*2-1}] $, where:

\begin{equation}
\resizebox{0.9\hsize}{!}{$
	a_i = \begin{cases}
	\#\{e \in In(v): i = \mathcal{M}_e(\mathcal{X}_e(e)) \}, & i \in \{0, ..., N_e-1\} \\
	\#\{e \in Out(v): i = \mathcal{M}_e(\mathcal{X}_e(e)) + N_e \}, & i \in \{N_e, ..., N_e*2-1\}
		   \end{cases}
$}
\end{equation}

Through our feature extraction method, the features of benign node \textcircled{2} and anomalous node \textcircled{1} in Figure \ref{fig:1} are [0,0,0,0,0,0,0,2,1,0,0,0,0,0] and [0,0,0,0,0,0,0,0,0,1,1,1,1,25301], which are significantly different in the last dimension.

\textbf{Training Method.} In the training phase, we split the whole training graph into several subgraphs. Each subgraph is constructed by a number (we set it as 150000 in this paper) of randomly selected $active$ nodes, their $related$ nodes, and edges between them. The sets of $active$ nodes in different subgraphs are disjointed. Then, the model is trained by each subgraph $G$ in order. Instead of using the whole graph for training, we only need to store one subgraph with a limited size in memory, which can guarantee scalability.

Our model is based on GraphSAGE, which uses a forward propagation (FP) algorithm \cite{23} to aggregate information from a node's ancestors. A GraphSAGE model uses graph information $G = (V, E, \mathcal{X}_v, \mathcal{X}_e, \mathcal{T}_e)$, features assigning function $ \mathcal{F}$, hop number $K$, neighborhood function $\mathcal{N} : V \rightarrow 2^V$, and parameters of GraphSAGE model \cite{23} (a set of weight matrices $\mathbf{W}^k, \forall k \in \{1, ..., K\}$; aggregator functions ${AGGREGATE}_k, \forall k \in \{1, ..., K\}$; nonlinear activation function $\sigma$; the batch mode parameter $Batch$ $size$) as input, and outputs the vector representations $z_v$ for $\forall v \in V$. The parameters (e.g. aggregator functions, which are \textit{mean aggregator} in our detector) of GraphSAGE are used to aggregate and process information from a node's ancestors. $z_v$ is a $N_n$-dimensional vector and the index of the biggest element in $z_v$ means the predicted class. The choice of $Batch$ $size$ $(BS)$ affects the runtime overhead and we evaluate it in \S\ref{section:6.5}. $K$ denotes the size of ancestors from that each node aggregates information. We set $K$ as 2 in this paper to balance the representing ability and runtime overhead. This is also the reason of defining $related$ nodes as nodes that can reach $active$ nodes in two hops (\S\ref{section:4.2}). Note that we have $K-1$ hidden layers in our model because the aggregation happens between two layers: the first aggregation happens from the input layer to the hidden layer, and the second aggregation happens from the hidden layer to the output layer. The FP algorithm repeats several times. After each iteration, we calculate the cross-entropy loss of $z_v$ and tune the weight matrices $\mathbf{W}^k, \forall k \in \{1, ..., K\}$. Through the FP and tuning procedure, the model learns the representation of different classes of benign nodes by exploring their features and local structures. 

Under a stealthy intrusion campaign, the proportion of anomalous nodes in the provenance graph may be small. Therefore, it is more likely to raise false positives that may disturb the judgment (\hyperlink{C2}{\textbf{C2}}). Motivated by the characteristics of intrusion detection scenes, we tailor a multi-model framework to reduce false positives. In an intrusion detection scene: 1) The number of different types of nodes is very unbalanced (e.g., there may be thousands of $process$ nodes while only one $stdin$ node). This critical unbalanced characteristic causes difficulty for a single model to learn the representation of those nodes with a small proportion. 2) Nodes with the same type may have different missions (e.g., the \texttt{fluxbox} process \textcircled{2} and \texttt{bash} process \textcircled{3} in Figure \ref{fig:1}), which makes it hard to classify those nodes into one class by one model. Therefore, it is unrealistic to train only one model that can distinguish all classes of benign nodes.

\begin{algorithm}[t]
\setlength{\abovecaptionskip}{0.cm}
\setlength{\belowcaptionskip}{-0.3cm}
\caption{The training method of the multi-model framework}
\label{alg:3}
\LinesNumbered
\KwIn{Subgraph $G = (V, E, \mathcal{X}_v, \mathcal{X}_e, \mathcal{T}_e)$; Features mapping function $\mathcal{F} : V \rightarrow \mathbb{N}^{2*N_e}$; Label mapping function $\mathcal{L} : V \rightarrow \{0, ..., N_n\}$; Hop number $K$; Times that FP algorithm iterates $epoch$}
\KwOut{ Number of submodels $cnt$; Submodels \{$\mathbb{M}_0, \mathbb{M}_1, ..., \mathbb{M}_{cnt-1} $\}}
Remove the correctly classified $active$ nodes from $V$\;
$X \leftarrow active$ nodes in $V$\;
$ cnt \leftarrow 0 $\; 
\While{$X$ not empty}{
	$\hat{V} \leftarrow \text{nodes in } K \text{-hop neighborhood of } \forall v \in X $\;
	\For {$k = 1 ... \text{epoch}$} {
		$\hat{G} \leftarrow (\hat{V}, E, \mathcal{X}_v, \mathcal{X}_e, \mathcal{T}_e) $ \;
		$z \leftarrow \text{FP}(\hat{G}, \mathcal{F}, K, AGGREGATE_k, \mathbf{W}^k, \sigma)$\;//$z$ are the representations of $\forall v \in \hat{V}$

		$\hat{z} \leftarrow z_{\forall v \in X}$\;
		$loss \leftarrow \text{cross\_entropy}(\hat{z}, \mathcal{L}(X))$\;
		$\mathbf{W}^k \leftarrow \text{backward}(loss)$\;
	}

$M \leftarrow \text{softmax}(\hat{z})$\;

\For {$v \in X$} {
$C_v \leftarrow$ index of the biggest element in $M_v$\;
$\hat{C}_v \leftarrow$ index of the second biggest element in $M_v$\; 

\If {$C_v = \mathcal{L}(v) \text{ and } M_{vC_v}/M_{v\hat{C_v}} > R_t$} {
	remove $v$ from $X$ \;
}

}
$\mathbb{M}_{cnt} \leftarrow$ current trained submodel\;
$cnt \leftarrow cnt + 1$

}
\vspace{-1mm}
\end{algorithm}

We consider a node $v$ has a dominant label $\mathcal{L}(v) = \mathcal{M}_v( \mathcal{X}_v(v))$ and a hidden label $\hat{\mathcal{L}}(v)$. The hidden label represents its specific function, which is just a concept for distinguishing nodes with the same dominant label and we do not know its value. We train multiple submodels to learn the representation of those nodes with small proportions and those nodes with the same dominant label but different hidden labels. The training method for the multi-model framework is shown in Alg. \ref{alg:3}. We maintain a list $X$, which stores the nodes that have not been correctly classified. It is initialized with all $active$ nodes in the training subgraph. After training a submodel (line 6-13), we validate nodes of $X$ in this model (line 15-21) and delete the correctly classified nodes from $X$. The nodes left in $X$ will be used to train a new submodel. We repeat this procedure until no nodes are left in $X$, which means that every node in the subgraph has been used to train a submodel that learns its representation. 

We design a probability-based method to further reduce the risk of false positives (line 18-20). By validating nodes of $X$ using a trained submodel, we receive a $|X| * N_n$ probabilistic matric $M$ where:

\vspace{-0.3cm} 
\begin{equation}
M_{ij} = P(\mathcal{L}(X_i) = j)
\end{equation}
\vspace{-0.3cm} 

For every node $X_i \in X$, we mark $C_i$ and $\hat{C_i}$ as the index of the biggest element and second biggest element in $M_i$. The naive method considers node $X_i$ has been correctly classified when $C_i = \mathcal{L}(X_i)$. However, consider this situation: $M_{iC_i}$ is not significantly bigger than $M_{i\hat{C_i}}$, which means that this submodel does not have a high degree of confidence in correctly classifying the node $X_i$. Thus, we set a threshold $R$ and consider $X_i$ is correctly classified by this submodel when:

\vspace{-0.3cm} 
\begin{equation}
C_i = \mathcal{L}(X_i) \text{ and } \frac{M_{iC_i}}{M_{i\hat{C_i}}} > R_t
\end{equation}
\vspace{-0.2cm}
 
During the training phase, \textsc{threaTrace} produces more submodels with a higher $R$, which thus results in fewer false positives during executing. By designing the multi-model framework and the probability-based method, \textsc{threaTrace} achieves a lower false positive rate (\hyperlink{C2}{\textbf{C2}}). The evaluation is presented in \S\ref{section:6.4}.

Note that because we split the whole training graph into several subgraphs, we need to carry out Alg. \ref{alg:3} several times, which may result in too many submodels. Therefore, we apply a heuristic method (line 1). We validate the subgraph $G$ in the existing submodels and remove the correctly classified $active$ nodes from $V$. It is based on the insight that the multi-model framework has already learned the representations of those removed nodes. By applying the pre-filtering strategy, we reduce the number of initialized $active$ nodes in $X$, which further reduces the number of submodels. More macroscopically, we apply a similar heuristic method for graphs in the training set. We use a graph for training only when it has nodes that the current multi-model framework cannot correctly classify. It is incremental work, which prevents producing ``excess" submodels.

By designing the multi-model framework and the probability-based method, \textsc{threaTrace} achieves a lower false positive rate (\hyperlink{C2}{\textbf{C2}}). The evaluation is presented in \S\ref{section:6.4}.

\textbf{Executing Method.} In the executing phase, we maintain a subgraph $\hat{G}$ in memory. $\hat{G}$ is empty originally and keeps appending with newly-arrived nodes and edges. The incoming edges' destination nodes and their 2-hop descendants are defined as $active$ nodes. The descendants and $related$ nodes are obtained from the whole graph in disk. We set a parameter called $Subgraph$ $Size$ $(SS)$. When the number of newly-arrived edges reaches $SS$, we use $\hat{G}$ for detection and start constructing a new subgraph in memory. Note that $SS$ is not the number of edges in $\hat{G}$ because we additionally add some edges from the graph in disk. Through the streaming executing mode, we can dynamically detect the currently active entities of the system and guarantee long-term scalability.

We design a multi-model framework and a probability-based training method to reduce false positives. However, there is another challenge of false negatives (\hyperlink{C3}{\textbf{C3}}). To tackle it, we also propose a probability-based method in the executing phase. For an abnormal node $v$, which is under detected, we detect it iteratively in submodels and get the probability list $M_v$ of each submodel. We mark $C_v$ and $\hat{C_v}$ as the class with the biggest and second biggest probability node $v$ to be. Because $v$ is an abnormal node with different behavior from benign nodes, it may not be classified to any class with high probability in every submodel. However, there exists a risk that it may still be correctly classified to its dominant label $\mathcal{L}(v)$ in one of the submodels just because the probability of other classes in this submodel is lower. Once it is correctly classified, the detector raises a false negative. Thus, during the executing phase, we also use the threshold $R$ and consider node $v$ is correctly classified only when:

\vspace{-0.3cm} 
\begin{equation}
\label{equa:4}
C_v = \mathcal{L}(v) \text{ and } \frac{M_{vC_v}}{M_{v\hat{C_v}}} > R_t
\end{equation}

\vspace{-0.1cm} 

It means that a submodel classifies node $v$ to the correct class with high assurance when equation \ref{equa:4} is satisfied. If $v$ does not satisfy equation \ref{equa:4} in every submodel, we consider it an abnormal node. For anomalous nodes during executing, a higher $R$ means the detector has a stricter standard of considering them as benign, which reduces false negatives (\hyperlink{C3}{\textbf{C3}}). For benign nodes, because more submodels for classifying benign nodes are trained with a higher $R$, the increase of false positives is less than the decrease of false negatives. Note that it does not mean $R$ should be set as high as possible because the model is difficult to converge when $R$ is too high. We further discuss the selection of $R$ detailedly in \S\ref{section:6.4}. Besides, although we detect nodes in $\hat{V}$ iteratively in every submodel, the size of $\hat{V}$ is reducing. Therefore, the number of submodels does not have a significant impact on processing speed. We evaluate the processing speed of \textsc{threaTrace} in \S\ref{section:6.5}.

\textbf{The final prediction of the multi-model framework.} A node is detected as benign if it is correctly classified in at least one submodel. Otherwise, it will be detected as anomalous.

\subsection {Alert and Trace}

\label{section:4.4}

This component receives abnormal nodes from the previous component and stores them in a queue $Q$. We set a time threshold $T$ and maintain $Q$ dynamically instead of directly raising alerts to the monitored system. It is because \textsc{threaTrace} detects abnormal nodes in a streaming mode to detect threats in an early stage while learning nodes' roles by the whole graph in the training stage in order to learn the rich contextual information. Therefore, a benign node may be detected as abnormal before it reaches its final stage. We first store abnormal nodes in $Q$ and set a waiting time threshold $T$ for it to reach its final stage. If it has not changed to benign within time $T$, we pop it from $Q$ and consider it as an abnormal node. A node may be detected several times under the streaming mode. Therefore, an anomalous node may be detected as benign at first and when it starts performing the malicious activity, it will be detected again as an $active$ node. When the number of abnormal nodes exceeds a tolerant threshold $\hat{T}$, which denotes the maximum number of false positives that may be generated, \textsc{threaTrace} raises an alert for the system and traces anomaly in the 2-hop ancestors and descendants of abnormal nodes. We discuss the selection of $T$ and $\hat{T}$ in \S\ref{section:6.4}.

%% file: evaluation.tex
\section{Evaluation}

\label{section:6}

We evaluate \textsc{threaTrace} using three public datasets, which are frequently used by state-of-the-art host-based threats detector evaluation. We focus on the following aspects:

\begin{itemize}[leftmargin=7mm]

\item[\textbf{Q1.}] The threats detection performance compared to state-of-the-art detectors. (\S\ref{section:6.1}, \ref{section:6.2}, \ref{section:6.3})

\item[\textbf{Q2.}] The influence of parameters. (\S\ref{section:6.4})

\item[\textbf{Q3.}] The ability to trace the abnormal behavior. (\S\ref{section:6.3})

\item[\textbf{Q4.}] The ability to detect abnormal behavior in the early stage of an intusion. (\S\ref{section:6.3})

\item[\textbf{Q5.}] The runtime and system resource overhead. (\S\ref{section:6.5})

\item[\textbf{Q6.}] The robustness against adaptive attacks. (\S\ref{section:6.6})

\end{itemize}

\newcommand{\tabincell}[2]{\begin{tabular}{@{}#1@{}}#2\end{tabular}}
\begin{table}[b]
\vspace{-2mm}
\setlength{\abovecaptionskip}{0cm}
\setlength{\belowcaptionskip}{-0.3cm}
\centering
\caption{Parameters of experiments.}
\scalebox{0.70}{
\begin{tabular}{|c|c|c|c|}
\hline
\textbf{Parameter}&\textbf{Description}&\textbf{Section}&\textbf{Values} \\
\hline
\text{$BS$}&\tabincell{c}{Batch mode parameter of GraphSAGE \cite{23}}&\text{\ref{section:4.3}}&\text{$5000$} \\
\hline
\text{$SS$}&\tabincell{c}{Maximun number of $active$ nodes}&\text{\ref{section:4.3}}&\text{$200000$} \\
\hline
\text{$R$}&\tabincell{c}{Probability rate threshold}&\text{\ref{section:4.3}}&\text{$1.5$} \\
\hline
\text{$T$}&\tabincell{c}{Waiting time threshold}&\text{\ref{section:4.4}}&\text{$168$} \\
\hline
\text{$\hat{T}$}&\tabincell{c}{Tolerant threshold}&\text{\ref{section:4.4}}&\text{$2$} \\
\hline
\end{tabular}
}
\label{table:1}
\vspace{-1mm}
\end{table}

\begin{table}[b]
\setlength{\belowcaptionskip}{-0.2cm}
\centering

\caption{Overview of StreamSpot dataset.}
\label{table:1.5}
\scalebox{0.9}{
\begin{tabular}{|c|c|c|c|}
\hline
\textbf{Scene}&\textbf{\# of graph}&\textbf{Average \# of nodes}&\textbf{Average \# of edges} \\
\hline
\text{Benign} & \text{500} & \text{8315} & \text{173857}\\
\hline
\text{Attack} & \text{100} & \text{8891} & \text{28423}\\
\hline
\end{tabular}
}
\vspace{0mm}
\end{table}

\begin{table}[b]
\setlength{\abovecaptionskip}{0.cm}
\setlength{\belowcaptionskip}{-0.3cm}
\centering

\caption{Overview of Unicorn SC-2 dataset.}
\label{table:2.5}
\scalebox{0.9}{
\begin{tabular}{|c|c|c|c|}
\hline
\textbf{Scene}&\textbf{\# of graph}&\textbf{Average \# of nodes}&\textbf{Average \# of edges} \\
\hline
\text{Benign} & \text{125} & \text{238338} & \text{911153}\\
\hline
\text{Attack} & \text{25} & \text{243658} & \text{949887}\\
\hline

\end{tabular}
}
\vspace{0mm}
\end{table}

\begin{table}[b]
\setlength{\abovecaptionskip}{0.cm}
\setlength{\belowcaptionskip}{-0.3cm}
\centering
\caption{Overview of DARPA TC dataset.}
\label{table:4}
\scalebox{0.8}{
\begin{tabular}{|c|c|c|c|c|}
\hline
\textbf{Scene}&\textbf{System}&\textbf{\# of benign nodes}&\textbf{\# of abnormal nodes}&\textbf{\# of edges}\\
\hline
\text{THEIA} & \text{Ubuntu} & \text{3505326} & \text{25362} & \text{102929710}\\
\hline
\text{Trace} & \text{Ubuntu} & \text{2416007} & \text{67383} & \text{6978024}\\
\hline
\text{CADETS} & \text{FreeBSD} & \text{706966} & \text{12852} & \text{8663569}\\
\hline
\text{fivedirections} & \text{Windows} & \text{569848} & \text{425} & \text{9852465}\\
\hline
\end{tabular}
}
\vspace{0mm}
\end{table}

Here we introduce the dataset, experimental setup, and implementation of comparison work. In \S\ref{section:6.1}, \ref{section:6.2}, we compare \textsc{threaTrace} to Unicorn, ProvDetector and StreamSpot, three state-of-the-art anomaly-based host threats detectors using their own datasets: StreamSpot and Unicorn SC-2 datasets. For StreamSpot and Unicorn, we implement them with the open-source projects. The results of them are almost the same as the original papers. For ProvDetector, we reimplement the approaches based on the original paper. For ProvDetector, we cannot compare the performance with the original paper because we do not have the private dataset of the paper for evaluation. We do not compare \textsc{threaTrace} to other related state-of-the-art detectors because: 1) Misuse-based detectors require a priori expert knowledge to construct a model \cite{7,8,58,65}; 2) The source input is not data provenance \cite{52,55,59,65}. These approaches are introduced in \S\ref{section:8}. 

We show that \textsc{threaTrace} can achieve better detection performance (\textbf{Q1}) in both datasets. In \S\ref{section:6.4}, we evaluate how the specially designed parameters influence the threats detection performance (\textbf{Q2}). In \S\ref{section:6.3}, we use the DARPA TC dataset to further evaluate \textsc{threaTrace}'s capability to trace anomaly (\textbf{Q3}) and detect persistent intrusions in the early stage (\textbf{Q4}). We evaluate \textsc{threaTrace}'s runtime performances in terms of executing speed and system resource overhead (\textbf{Q5}) in \S\ref{section:6.5}. We evaluate the robustness against adaptive attacks (\textbf{Q6}) in \S\ref{section:6.6}. Table \ref{table:2}, \ref{table:3}, \ref{table:4} show the overview of the experimental datasets. 

The manually set parameters of experiments in \S\ref{section:6.1}, \ref{section:6.2}, \ref{section:6.3}, \ref{section:6.6} are shown in Table \ref{table:1}. We set these parameters based on the detection and runtime performance in experiments of each subsection. Each submodel has the same GraphSAGE's hyperparameters. Specifically, we set most of those hyperparameters as default values and set the number of hidden layer's neurons as 32, which is approximately half of the features' number in Unicorn SC-2 dataset. The submodels have one hidden layer because we set $K = 2$, which denotes ancestors' size from which each node aggregates information. We test the influence of parameters in \S\ref{section:6.4}.

\begin{table}[b]
\setlength{\abovecaptionskip}{0.cm}
\setlength{\belowcaptionskip}{-0.3cm}
\centering

\caption{Comparison to StreamSpot and Unicorn on the StreamSpot dataset.}
\label{table:2}
\scalebox{0.65}{
\begin{tabular}{|c|c|c|c|c|c|c|c|c|c|}
\hline
\textbf{Detector}&\textbf{Precision}&\textbf{Recall}&\textbf{Accuracy}&\textbf{F-Score}&\textbf{TP}&\textbf{TN}&\textbf{FP}&\textbf{FN}&\textbf{FPR} \\
\hline
\text{StreamSpot} & \text{0.72} & \text{1.0} & \text{0.69}& \text{0.75}&  \text{25}&  \text{108.5}&  \text{16.5}&  \text{0}&  \text{0.11}\\
\hline
\text{Unicorn} & \text{0.95} & \text{0.97} & \text{0.99}& \text{0.96}& \text{24.32}& \text{123.64}& \text{1.36}& \text{0.68}& \text{0.011}\\
\hline
\text{\textsc{threaTrace}} & \text{1.0} & \text{1.0} & \text{1.0}& \text{1.0}& \text{25}& \text{125}& \text{0}& \text{0}& \text{0}\\
\hline
\end{tabular}
}
\vspace{0mm}
\end{table}

\subsection{StreamSpot Dataset}

\label{section:6.1}

We compare \textsc{threaTrace} to two state-of-the-art detectors: StreamSpot and Unicorn. 

\textbf{Dataset}. The StreamSpot dataset (Table \ref{table:1.5}) is StreamSpot's own dataset which is publicly available \cite{32}. It contains 6*100 information flow graphs derived from five benign scenes and one attack scene. We use the same validation strategy as Unicorn to randomly split the dataset into a training set with 75*5 benign graphs and a testing set with 25*5 benign graphs and 25 attack graphs. We do not compare with ProvDetector because edges in StreamSpot dataset do not have the timestamp attribute, which is necessary for ProvDetector. We repeat this procedure and report the mean evaluation results. 

\textbf{Result}. As shown in Table \ref{table:2}, \textsc{threaTrace} can achieve a perfect detection performance in this dataset. We analyze deep into this dataset and find that the 100 graphs of the attack scene can be generally split into two classes with 95 and 5 graphs. The 5 graphs in the second group are significantly different from the other 95 graphs. They are more similar to the graph of the Youtube scene, which means that threats in the second group are more stealthy. Drive-by download attack happened when the victim browses a website, which is similar to the youtube scene. It is hard for a graph-kernel-based method to detect those 5 graphs, which may explain why Unicorn has a high detection performance in this dataset but still cannot detect every abnormal graph. This result demonstrates \textsc{threaTrace}'s ability to detect the stealthy intrusion campaign.

\begin{table}[b]
\setlength{\abovecaptionskip}{0.cm}
\setlength{\belowcaptionskip}{-0.3cm}
\centering

\caption{Comparison to Unicorn and ProvDetector on the Unicorn SC-2 dataset.}
\label{table:3}
\scalebox{0.62}{
\begin{tabular}{|c|c|c|c|c|c|c|c|c|c|}
\hline
\textbf{Detector}&\textbf{Precision}&\textbf{Recall}&\textbf{Accuracy}&\textbf{F-Score}&\textbf{TP}&\textbf{TN}&\textbf{FP}&\textbf{FN}&\textbf{FPR} \\
\hline
\text{Unicorn} & \text{0.75} & \text{0.80} & \text{0.77}& \text{0.78}& \text{20}& \text{18.33}& \text{6.67}& \text{5}& \text{0.27}\\
\hline
\text{ProvDetector} & \text{0.67} & \text{0.6} & \text{0.65}& \text{0.63}& \text{15}& \text{17.5}& \text{7.5}& \text{10}& \text{0.3}\\
\hline
\textsc{threaTrace} (K = 1) & \text{0.81} & \text{0.79} & \text{0.8}& \text{0.8}& \text{19.7}& \text{20.5}& \text{4.5}& \text{5.3}& \text{0.18}\\
\hline
\textsc{threaTrace} (K = 2) & \text{0.91} & \text{0.96} & \text{0.93}& \text{0.93}& \text{24}& \text{22.5}& \text{2.5}& \text{1}& \text{0.1}\\
\hline
\end{tabular}
}

\vspace{0mm}
\end{table}

\subsection{Unicorn SC-2 Dataset}

\label{section:6.2}

This dataset is Unicorn's own dataset \cite{49}, which is more complex than the StreamSpot dataset. We use it to compare \textsc{threaTrace}'s detection performance with Unicorn and ProvDetector. We cannot use StreamSpot for comparison because it cannot deal with a large number of edges \cite{19}. 

\textbf{Dataset}. We follow the same 5-fold cross-validation as Unicorn: use 4 groups (each group contains 25 graphs) of benign graphs to train, and the 5th group of benign graphs and 25 attack graphs for validation. We use a streaming mode to replay the validation graphs and detect them dynamically. For \textsc{threaTrace}, we evaluate the detection performance with different neighborhood size, which is setted as 1 and 2 (default). We repeat this procedure and report the mean evaluation results.
	
 \textbf{Result}. As shown in Table \ref{table:3}, \textsc{threaTrace} achieves a better performance when $K = 2$, which indicates that adequate neighbor information learning is helpful for detection. We further discuss the setting of $K$ in \S\ref{section:7}. \textsc{threaTrace} can achieve better detection performance than Unicorn using its own dataset. The reason that Unicorn has a low performance is introduced in Unicorn's paper \cite{19}: the attacker has prior knowledge about the system and thus acts more stealthily than attackers of other datasets. We also find that \textsc{threaTrace} raises alerts for less than 10 nodes in most attack graphs, which demonstrates the stealthy characteristic of this dataset. The comparison results in this dataset further demonstrate our motivation that the graph-kernel-based method is hard to detect stealthy threats. Besides, \textsc{threaTrace} also raises fewer false positives than Unicorn and ProvDetector, which is important to alleviate the ``threat fatigue problem".

\subsection{DARPA TC Dataset}

\label{section:6.3}

\begin{figure}[htbp]
\setlength{\abovecaptionskip}{0.cm}
\setlength{\belowcaptionskip}{-0.3cm}
\centerline{\includegraphics[width=0.45\textwidth]{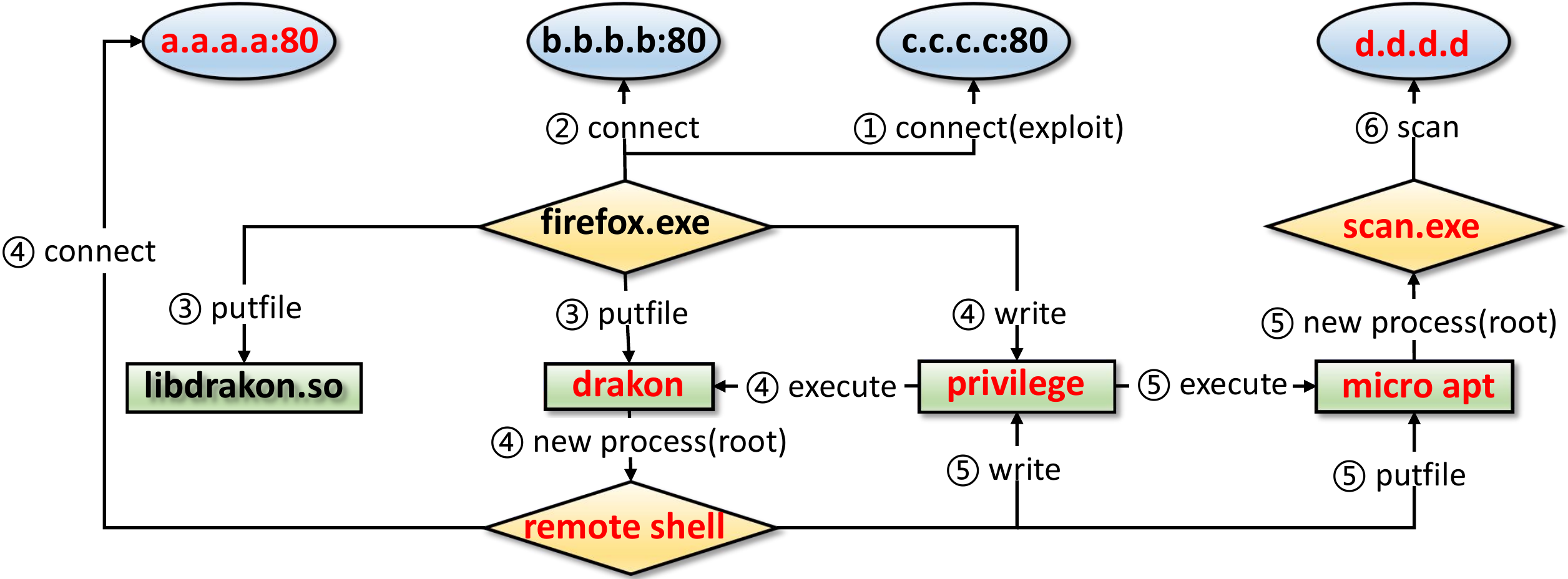}}
\caption{The case study. Entities in red are those detected by \textsc{threaTrace}.}
\label{fig:6}
\vspace{-2mm}
\end{figure}

\begin{table}[b]
\setlength{\abovecaptionskip}{0.cm}
\setlength{\belowcaptionskip}{-0.3cm}
\centering

\caption{Results of node level detection in DARPA TC dataset}
\label{table:5}
\scalebox{0.65}{
\begin{tabular}{|c|c|c|c|c|c|c|c|c|c|}
\hline
\textbf{Experiment}&\textbf{System}&\textbf{Precision}&\textbf{Recall}&\textbf{F-Score}&\textbf{TP}&\textbf{TN}&\textbf{FP}&\textbf{FN}&\textbf{FPR}\\
\hline
\text{THEIA} & \text{Ubuntu} & \text{0.87} & \text{0.99}  & \text{0.93}& \text{25297}& \text{3501561}& \text{3765}& \text{65}& \text{0.001}\\
\hline
\text{Trace} & \text{Ubuntu}  & \text{0.72} & \text{0.99} & \text{0.83}& \text{67382}& \text{2389233}& \text{26774}& \text{1}& \text{0.011}\\
\hline
\text{CADETS} & \text{FreeBSD} & \text{0.90} & \text{0.99} & \text{0.94}& \text{12848}& \text{705605}& \text{1361}& \text{4}& \text{0.002}\\
\hline
\text{fivedirections} & \text{Windows} & \text{0.67} & \text{0.92} & \text{0.80}& \text{389}& \text{569660}& \text{188}& \text{36}& \text{0.0003}\\
\hline
\end{tabular}
}
\vspace{0mm}
\end{table}

We use DARPA TC dataset to evaluate \textsc{threaTrace}'s ability to detect and trace long-term running intrusions. Datasets in the last two subsections do not have the ground truth of nodes and thus, we cannot use them to evaluate \textsc{threaTrace}'s ability of anomaly tracing.

\textbf{Dataset}. DARPA TC dataset (Table \ref{table:4}) is generated in the third red-team vs. blue-team engagement of the DARPA Transparent Computing program. The engagement lasted for two weeks and the provenance collecting tools captured provenance data of the whole system from start to end. The provenance data and ground truth are publicly available \cite{47}. Therefore, we use the ground truth to label the abnormal nodes and evaluate \textsc{threaTrace}'s performance of detecting them, which shows the ability of tracking anomaly. We remove part of the dataset because some graphs are generated in exceptional accidents such as outages and shutdown of hosts. We use nodes in benign graphs to train our models and use nodes in graphs containing threats for evaluation. Because we check a node's 2-hop ancestors and descendants to track its position (\S\ref{section:4.4}) when it is detected as an anomaly, we define metrics as follows:

\begin{itemize}[leftmargin=3mm]

\item \textbf{True Positive.} The anomalous nodes which are detected as abnormal or the anomalous nodes one of whose 2-hop ancestors and descendants has been detected as abnormal. Those nodes are defined as true positives because we can hunt them during alert tracing. 

\item \textbf{False Positive.} The benign nodes which are detected as abnormal and do not have anomalous nodes in their 2-hop ancestors and descendants. Those nodes are defined as false positives because we cannot hunt any anomalous nodes during alert tracing of them.

\item \textbf{True Negative.} Other benign nodes.

\item \textbf{False Negative.} Other anomalous nodes.

\end{itemize}

We assume the correctness of the manual inspection of nodes' 2-hop ancestors and descendants. The workforce cost of tracing anomaly is acceptable since the average number of nodes' 2-hop ancestors and descendants is $1.8$ in this dataset. Because a node in a provenance graph denotes an entity in a system, it is not difficult to inspect its behavior manually. 

\begin{table}[b]

\vspace{0mm}
\setlength{\abovecaptionskip}{0.cm}
\setlength{\belowcaptionskip}{-0.3cm}
\centering

\caption{Results of graph level detection in DARPA TC dataset}
\label{table:6.3}
\scalebox{0.7}{
\begin{tabular}{|c|c|c|c|c|c|c|c|}
\hline
\textbf{Experiment}&\textbf{Detector}&\textbf{Precision}&\textbf{Recall}&\textbf{F-Score}&\textbf{FP}&\textbf{FN}&\textbf{FPR}\\
\hline
\multirow{3}{*}{THEIA} & \text{\textsc{threaTrace}} & \text{1.0} & \text{1.0}  & \text{1.0}& \text{0}& \text{0}& \text{0}\\
\cline{2-8}
 & \text{Unicorn} & \text{1.0} & \text{1.0}  & \text{1.0}& \text{0}& \text{0}& \text{0}\\
\cline{2-8}
 & \text{ProvDetector} & \text{0.96} & \text{0.92}  & \text{0.94}& \text{0.2}& \text{0.4}& \text{0.04}\\

\hline

\multirow{3}{*}{CADETS} & \text{\textsc{threaTrace}} & \text{1.0} & \text{1.0}  & \text{1.0}& \text{0}& \text{0}& \text{0}\\
\cline{2-8}
 & \text{Unicorn} & \text{1.0} & \text{1.0}  & \text{1.0}& \text{0}& \text{0}& \text{0}\\
\cline{2-8}
 & \text{ProvDetector} & \text{0.92} & \text{0.94}  & \text{0.93}& \text{0.4}& \text{0.3}& \text{0.08}\\

\hline

\multirow{2}{*}{fivedirections} & \text{\textsc{threaTrace}} & \text{1.0} & \text{1.0}  & \text{1.0}& \text{0}& \text{0}& \text{0}\\
\cline{2-8}
 & \text{Unicorn} & \text{1.0} & \text{1.0}  & \text{1.0}& \text{0}& \text{0}& \text{0}\\
\cline{2-8}
 & \text{ProvDetector} & \text{1.0} & \text{1.0}  & \text{1.0}& \text{0}& \text{0}& \text{0}\\

\hline
\multirow{3}{*}{Trace} & \text{\textsc{threaTrace}} & \text{1.0} & \text{1.0}  & \text{1.0}& \text{0}& \text{0}& \text{0}\\
\cline{2-8}
 & \text{Unicorn} & \text{1.0} & \text{1.0}  & \text{1.0}& \text{0}& \text{0}& \text{0}\\
\cline{2-8}
 & \text{ProvDetector} & \text{1.0} & \text{1.0}  & \text{1.0}& \text{0}& \text{0}& \text{0}\\

\hline
\end{tabular}
}
\vspace{-6mm}

\end{table}

\textbf{Result}. The results in Table \ref{table:5} show \textsc{threaTrace}'s performance of detecting abnormal nodes, which means the ability of anomaly tracing. The results are acceptable in consideration of the highly imbalanced characteristic of this dataset. We cannot compare \textsc{threaTrace} with StreamSpot, Unicorn, and ProvDetector at node level because they do not make anomalous nodes detection. The detection results of graph level experiments are shown in Table \ref{table:6.3}. The results show that both \textsc{threaTrace} and the comparison methods have good detection performance. One explanation is that the attackers in the DARPA datasets spend time finding vulnerabilities that leave apparent behavior in the graphs. 

\textbf{Case study}. We use the motivation example (Figure \ref{fig:1}) as the case study to further describe \textsc{threaTrace}'s anomaly tracing ability. It is simplified as an attack graph in Figure \ref{fig:6}. The attack lasts for two days and the victim is a \texttt{Ubuntu 12.04 x64} host. The attacker exploits \texttt{Firefox 54.0.1} backdoor using a malicious website \texttt{c.c.c.c:80}. When the victim connects to this website (\textcircled{1}), a drakon implant begins running in memory and results in a connection to the attacker operator console \texttt{b.b.b.b:80} (\textcircled{2}). The drakon implant is not shown in the figure because there is not enough information about it in the ground truth file. The attacker further writes \texttt{libdarkon.so} and \texttt{drakon} executable binary to the victim host's disk (\textcircled{3}). Then, the attacker uses a \texttt{privilege} escalated execution capability to execute the \texttt{drakon}, which results in a process \texttt{remote shell} with root privilege (\textcircled{4}). The new \texttt{remote shell} process connects to a new operator console \texttt{a.a.a.a:80} with root access (\textcircled{4}) and the attacker stops their campaign temporarily. Two days later, the attacker writes a \texttt{micro apt} to disk and elevates it as a new process \texttt{scan.exe} with root privileges (\textcircled{5}). Finally, the attacker carries out a port scan of \texttt{d.d.d.d} (\textcircled{6}).

In this case, \textsc{threaTrace} raises alerts of the \texttt{darkon} executable binary, \texttt{remote shell}, \texttt{privilege} escalated execution, malicious ip \texttt{a.a.a.a:80}, \texttt{micro apt} and its process \texttt{scan.exe}, the scanned ip object \texttt{d.d.d.d}. \textsc{threaTrace} achieves this result in the situation that the anomalous nodes only take a small part (less than 1\%) in the whole provenance graph due to the deliberate prolonging of the attack duration, which shows \textsc{threaTrace}'s ability to detect and trace anomaly in a long-term running system. Besides, \textsc{threaTrace} successfully detects some early steps of the attack such as \texttt{darkon} and \texttt{privilege} which shows \textsc{threaTrace}'s ability to detect and prevent the intrusion in the early phases. We also note that \textsc{threaTrace} fails to detect some components of the attack like \texttt{libdarkon.so} and malicous ip addresses \texttt{b.b.b.b:80} and \texttt{c.c.c.c:80}. We find that these components do not show significant influence during the attack period: they act more like auxiliary roles. This partially explains \textsc{threaTrace}'s failure on detecting those attack parts. Fortunately, there is usually no need to detect every part of an intrusion campaign. In this case, \textsc{threaTrace} detects some important components such as the \texttt{darkon} executable binary, which is enough to detect and stop the intrusion in its early phase.

\begin{figure*}[htbp]
\setlength{\abovecaptionskip}{0.cm}
\setlength{\belowcaptionskip}{-0.3cm}
\centerline{\includegraphics[width=1.0\textwidth]{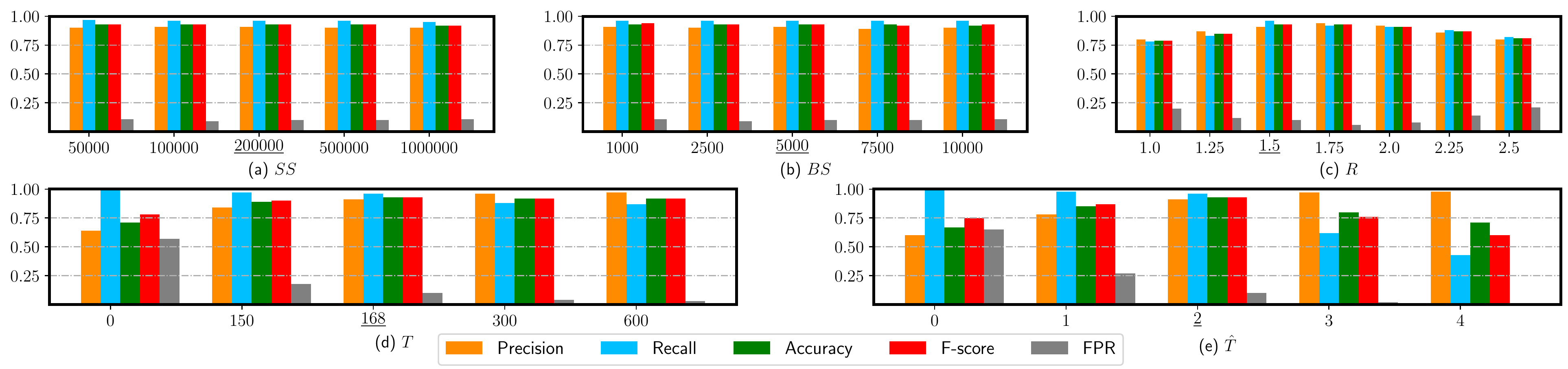}}
\caption{Detection performance with varying parameters. }
\label{fig:7}
\vspace{-2mm}
\end{figure*}

\subsection{Influence of Different Parameters}

\label{section:6.4}

\textsc{threaTrace} shows a detection performance improvement compared to the state-of-the-art detector in the Unicorn SC-2 dataset. We further use the Unicorn SC-2 dataset to study how the parameters, which are specially designed for host-based threats detection, influence the detection performance. The parameters we test are $SS$, $BS$, $R$, $T$, $\hat{T}$. The values in Tabel \ref{table:1} make up the baseline, which gains encouraging detection and runtime performance in Unicorn SC-2 dataset. When we test one of them, we keep the others the same as the baseline.

\begin{itemize}[leftmargin=*]

\item $Subgraph$ $Size$ $(SS)$. $SS$ is the parameter we set for the Data Storage module (\S\ref{section:4.2}). As shown in Figure \ref{fig:7}(a), it has little influence on detection performance. \textsc{threaTrace} detects the provenance graph in a streaming mode. The subgraph of current system execution is updated in the memory and is transferred to the detection module when its size reaches $SS$. Thus, $SS$ affects the coming interval of subgraphs, which may influence the time we set for benign nodes to reach their final stage. Besides, the randomly divided dataset may have an impact on the results as well. Compared to detection performance, $SS$ has much more influence on runtime performance (\S\ref{section:6.5}).

\item $Batch$ $Size$ $(BS)$. $BS$ is the parameter we set for the Model module (\S\ref{section:4.3}). As shown in Figure \ref{fig:7}(b), $BS$ has almost no influence on detection performance, which is probably caused by the randomly splitted dataset. We evaluate its influence on runtime performance in \S\ref{section:6.5}.

\item $R$. This is the rate threshold we set for the training phase. As shown in Figure \ref{fig:7}(c), the detection performance improves when $R$ increases from 1.0 to 1.5. $R$ has no significant influence on detection performance from 1.5 to 2.0. However, when it exceeds 2.0, the performance tends to decrease.

\item $T$ and $\hat{T}$. These two parameters influence the Alert and Trace module (\S\ref{section:4.4}). The results are shown in Figure \ref{fig:7}(d)(e), which can be explained by the function of $T$ and $\hat{T}$. $T$ is the waiting time threshold we set for benign nodes to reach their final stage. In real-time streaming detection mode, a benign node may be detected as an anomaly before it reaches its final stage. Thus, a bigger $T$ means we ``give more evolutionary time" to the benign nodes and results in higher $Precision$, lower $Recall$ and $FPR$. Note that although it decreases the number of false positives, it also decreases \textsc{threaTrace}'s ability to detect intrusion as early as possible: it needs more time to wait for an anomalous node before raising alert for it. The average time of raising an alert for an anomalous graph is 434.45 seconds when $T$ = 150, while it turns to 874.32 seconds when $T$ is increased to 600. $\hat{T}$ is the tolerant threshold for \textsc{threaTrace} to raise alerts for the monitored system. Bigger $\hat{T}$ means less false positives and more false negatives, which thus results in higher $Precision$, lower $Recall$ and $FPR$.

\end{itemize}

The evaluation of different parameters shows how \textsc{threaTrace}'s special design for host-based threats detection influences the detection performance. The results of each parameter can be explained by the design principle. We can leverage the function of different parameters to optimize \textsc{threaTrace}'s performance, which is discussed in \S\ref{section:7}.

\subsection{Runtime Overhead}

\label{section:6.5}

We study the processing speed, CPU utilization, and memory usage during \textsc{threaTrace}'s execution in Unicorn SC-2 dataset. This section focuses on \textsc{threaTrace}'s runtime performance in the execution phase. The experiments in this subsection are done in an Ubuntu 16.04.6 LTS machine with 16 vCPUs and 64GiB of memory. We reimplement Unicorn and ProvDetector and evaluate them in the same machine. We repeat the experiments and report the average results. 

\begin{figure}[htbp]

\vspace{-3mm}
\centerline{\includegraphics[width=0.5\textwidth]{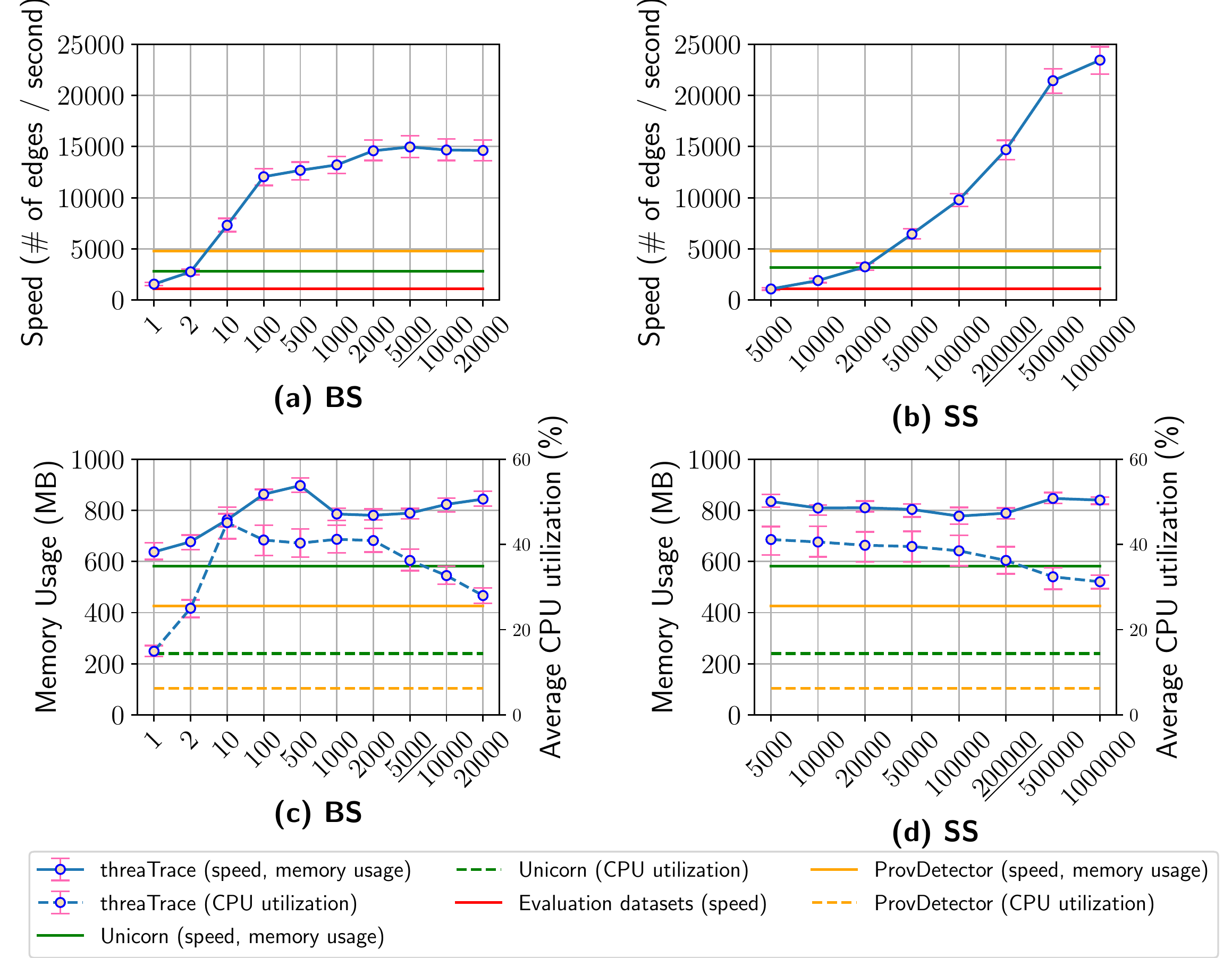}}
\caption{Runtime performance with varying parameters. The evaluation of each parameter is done with the remaining parameters constant.}
\label{fig:8}

\vspace{-5mm}
\end{figure}

\textbf{Processing speed}. We evaluate the processing speed of the entire detector, which is composed of four components. Figure \ref{fig:8}(a)(b) shows the processing speed of \textsc{threaTrace} in different parameters. We test \textsc{threaTrace}'s processing speed in different $Subgraph$ $Size$ $(SS)$ and $Batch$ $Size$ $(BS)$. The red line is the maximum edges generated speed ($M$) in the three datasets we previously discussed. The green and orange lines are the maximum processing speed of Unicorn and ProvDetector. The results show that even in the worse case, \textsc{threaTrace}'s processing speed is still faster than $M$. In the best case, \textsc{threaTrace}'s is about five and three times faster than Unicorn and ProvDetector. We test one parameter and set the others as the baseline in Table \ref{table:1}.

\begin{itemize}[leftmargin=*]

\item $Batch$ $Size$ $(BS)$. $BS$ is the parameter we set for the Model module (\S\ref{section:4.3}). As shown in Figure \ref{fig:8}(a), with the increase of $BS$, \textsc{threaTrace}'s processing speed has an increasing trend and tends to be flat.

\item $Subgraph$ $Size$ $(SS)$. $SS$ is the parameter we set for the Data Storage module (\S\ref{section:4.2}). As shown in Figure \ref{fig:8}(b), runtime performance improves as we increase $SS$.

\end{itemize}

\textbf{CPU utilization \& Memory usage}. Figure \ref{fig:8}(c)(d) demonstrates \textsc{threaTrace}'s CPU utilization and memory usage in different $SS$ and $BS$. The green and orange lines are the baseline results of Unicorn and ProvDetector. The results show that with the increase of $BS$, CPU utilization tends to be bigger and reaches its maximum value when $BS = 10$. After that, it becomes smaller. In terms of $SS$, CPU utilization has a decreasing trend when $SS$ increases. There is no significant impact of varying $BS$ and $SS$ on memory usage. \textsc{threaTrace}'s CPU utilization and memory usage are bigger than Unicorn and ProvDetector. This is primarily because \textsc{threaTrace} is a deep-learning-based method that needs more computation resources than traditional methods. It is a limitation of \textsc{threaTrace} and we plan to research the problem of decreasing computation resources in future work.

We do not present \textsc{threaTrace}'s runtime overhead with other parameters because they only influence detection performance. At the same time, $BS$ and $SS$ have little influence on detection performance. These independent and insensitive characteristics provide the convenience of tuning \textsc{threaTrace}.

\subsection{Adaptive attacks}

\label{section:6.6}

\begin{figure}[htbp]
\centerline{\includegraphics[width=0.45\textwidth]{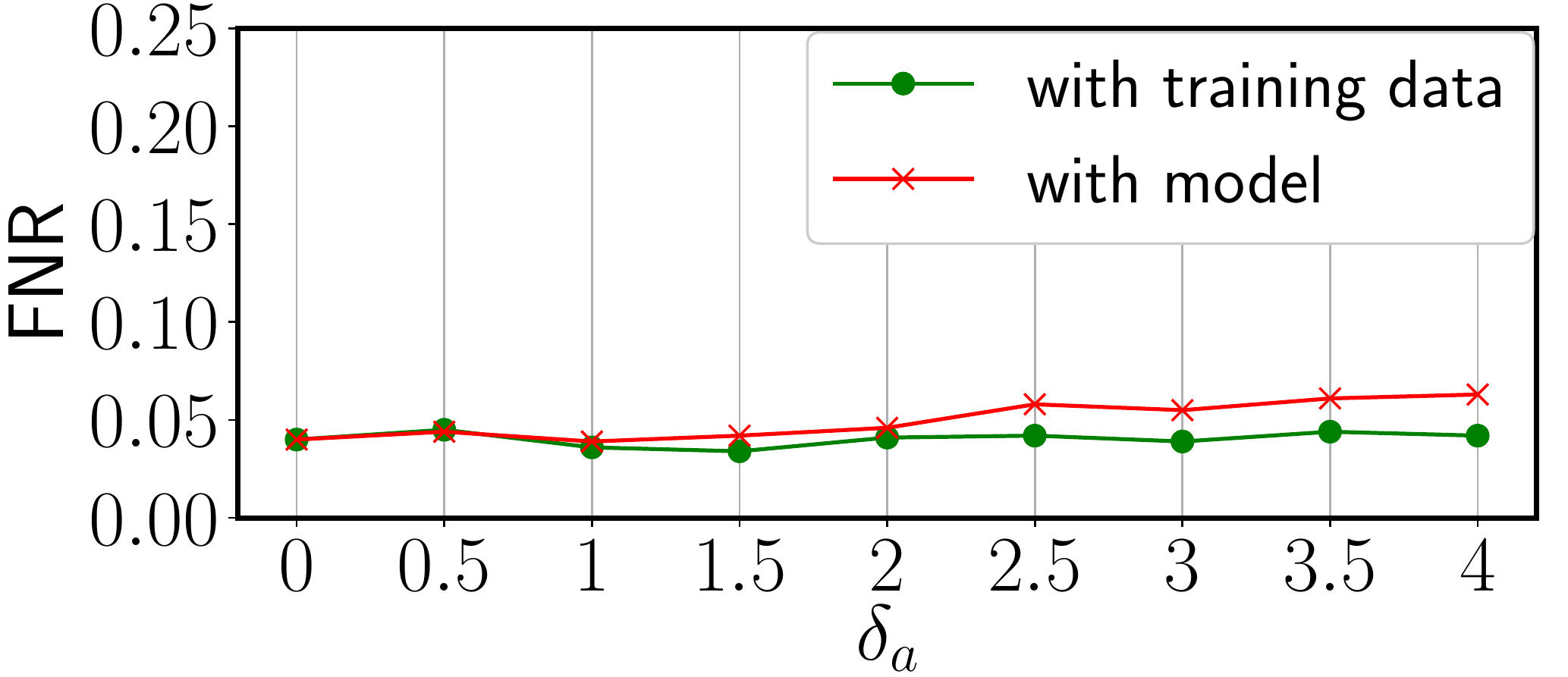}}
\caption{Result of evasion attack experiments.}
\label{fig:11}
\vspace{-5mm}
\end{figure}

\textbf{Attack Methodology.} \textsc{threaTrace} is a Graph-based detection system. Therefore, we borrow the idea of existing adversarial attack against Graph-based detectors \cite{67, 68, 69} to develop an adaptive attack on \textsc{threaTrace}, called optimization-based evasion attack. The purpose of this attack is to make abnormal nodes evade the detection of \textsc{threaTrace} in the execution phase. Specifically, because \textsc{threaTrace} judges the node as abnormal when it is misclassified, the purpose of the attacker is to make the abnormal node be classified into the correct class to avoid detection. The attacker needs to find a perturbation on the abnormal node's feature which is constructed with edges between the node and its neighbor. Note that we suppose the attacker has compromised the system by implanting some abnormal nodes (such as \textit{Malware}, \textit{Remote Shell}, anomalous \textit{Dynamic Link Library}) into it. Therefore, the attacker can control the related edges of the abnormal nodes. The perturbation should be as small as possible to keep the original function of the abnormal node and reduce cost. In a word, suppose $x$ is the feature of an abnormal node which can be detected by \textsc{threaTrace} originally, the optimization-based evasion attack's goal is to change the feature to $\hat{x}$ to evade detection with the constraint $\frac{\left\|\hat{x}-x\right\|_2}{\left\|x\right\|_2} < \delta_a$, where $\delta_a$ limits the perturbation. In order to construct a node's adversarial features $\hat{x}$, we assume that the attacker knows \textsc{threaTrace}'s feature extraction method mentioned in \S\ref{section:4}. For other adversarial's background knowledge of \textsc{threaTrace}, we study two kinds of attackers based on the background knowledge.

\textit{(1) Attackers with training data.} This kind of attacker has the training data. The optimization-based evasion attack with training data can be performed in two steps. The first step is to find a benign node $x_b$ in the training data, which is most similar to the anomalous node $x$ and has the same class as $x$. Formally,

\vspace{-0.3cm} 
\begin{equation}
\label{equa:6}
argmin_{x_b} \left\|x_b-x\right\|_2 \quad s.t. \quad class(x_b) = class(x)
\end{equation}
\vspace{-0.3cm} 

The second step is to solve the optimization problem, which is not difficult:

\vspace{-0.3cm} 
\begin{equation}
\label{equa:7}
argmin_{\hat{x}}\left\|\hat{x}-x_b\right\|_2 \quad s.t. \quad \frac{\left\|\hat{x}-x\right\|_2}{\left\|x\right\|_2} < \delta_a, \hat{x} \in \mathbb{N}
\end{equation}
\vspace{-0.3cm} 

\textit{(2) Attackers with \textsc{threaTrace}'s model.} This kind of attacker knows \textsc{threaTrace}'s model, including the parameters after training and hyperparameters. Therefore, the attacker can solve the optimization problem directly based on the loss of the model. Formally,

\vspace{-0.3cm} 
\begin{equation}
\label{equa:8}
argmin_{\hat{x}}(loss(\hat{x})) \quad s.t. \quad \frac{\left\|\hat{x}-x\right\|_2}{\left\|x\right\|_2} < \delta_a, \hat{x} \in \mathbb{N}
\end{equation}
\vspace{-0.3cm} 

$Loss(\hat{x})$ is the loss function of \textsc{threaTrace}, indicating whether a sample is classified into the correct category. For an anomalous sample, the attacker needs to make \textsc{threaTrace} correctly classify it to evade detection. Therefore, $loss(\hat{x})$ should be as small as better. This problem can be easily solved by a gradient-based method. \textsc{threaTrace} consists of several submodels. For simplicity, we choose the first submodel for the target model to evasion.

For these two kinds of attackers, once $\hat{x}$ is successfully solved, the attacker controls the abnormal node to interact with other nodes in the system according to the new distribution of edges, so that its feature will change to $\hat{x}$.

\noindent\textbf{Robustness Evaluation and Analysis.} We use Unicorn SC-2 dataset for evaluation. We first choose the abnormal nodes detected by \textsc{threaTrace} as the base samples and then apply the optimization-based evasion approach to compute $\hat{x}$ for each abnormal node. After that, we change the related edges of the adversarial samples to change their features extracted. The results are shown in Figure \ref{fig:11}. For attackers with training data, the imitation of training data has almost no effect on detection performance. For attackers with \textsc{threaTrace}'s model, the evasion effection increases with the increasing of $\delta_a$. However, compared to the original result, the FNR is acceptable (raise from 0.04 to 0.07), which demonstrates \textsc{threaTrace}'s good robustness against optimization-based evasion attack.

%% file: discussion.tex
\section{Discussion \& Limitation }

\label{section:7}

We discuss some issues, limitations, and future work of \textsc{threaTrace} in this section.

\textbf{Closed-world assumption}. Host-based threats detection is essentially a classification problem. It is based on a closed world assumption\cite{41, 42}, which assumes that every benign behavior has been involved in the training set and the rest are abnormal. However, in real-life problems, it is hard to guarantee that all benign cases are covered. The conflict between the closed-world assumption and real-life situations makes it hard to design an ideal detector. As the same as other anomaly-based detectors, \textsc{threaTrace} faces this problem too. We cannot guarantee that we learn every benign node in models especially considering the rapid development of present systems. However, system administrators can update models periodically using updated benign data provenance of the system. It is incremental work with no need to retrain the previous models and thus does not need much time (\S\ref{section:4.3}).

\textbf{Adversarial attacks}. The robustness goal of this paper is to be robust against the evasion attack that we study in \S\ref{section:6.6}. There are other adaptive attacks that may fail \textsc{threaTrace}, such as poisoning attack \cite{72} and graph backdoor \cite{71}. Future work can study the robustness against more attacks. Some pioneering methods for enhancing GNN's robustness can be found in \cite{53,54}. They are not suitable for \textsc{threaTrace} because they focus on robustness in graph convolutional networks, which are transductive models different from our inductive model (\S\ref{section:2.2}).

\textbf{Evaluation}. In host-based threats detection domain, public datasets are usually generated in a manually constructed environment \cite{32, 49} and thus may not represent typical workloads in practice. Data generated from red-team vs. blue-team engagement \cite{47} is partly closed to the actual situation. It is important to generate a public dataset for the community which can reflect the typical workloads in real situation and is in consideration of privacy and confidentiality. We plan to research this problem in future work. 

\textbf{System overhead.} It is necessary to maintain a stable system overhead during deployment. We demonstrate that \textsc{threaTrace} has a fast processing speed and acceptable system overhead in \S\ref{section:6.5}. However, if the host's performance is relatively low in practice, we need to adjust $BS$ and $SS$ to balance detection timeliness and system overhead. In the worst case, we may have to perform offline detection.

\textbf{Threat fatigue problem \cite{37}.} It is important to avoid ``threat fatigue problem'' by reducing false positives. Although we try to reduce false positives by multi-model framework and probability-based method, \textsc{threaTrace} still raises many false positives in DARPA TC datasets. It may be related to the large number of benign nodes (Table \ref{table:4}). \textsc{threaTrace} has limitations of controlling the number of false positives when there are plenty of nodes in the graph, and we seek means to alleviate this phenomenon. In practice, an enterprise or government usually has a list of reliable software. We suggest the administrator maintain a whitelist to record system entities that are related to those reliable softwares. Besides, if a false positive entity is inspected as benign and believed to be reliable in the future, it can also be added to the Whitelist. Whitelist is a typical method in recent work \cite{8, 58} for reducing false positives. Apart from the whitelist, the administrator can use false positives to train more submodels and thus avoid raising the same false positives in the future. 

\textbf{Unsupervised.} \textsc{threaTrace} needs benign data for training. In practice, the administrator can collect provenance data when the system is running without threats (e.g., only trusted software and websites can be used and visited).

\textbf{Granularity of data provenance}. \textsc{threaTrace} falls into the category of provenance-based methods. Some threats (e.g., malicious code in a file and thread-based threats) will not demonstrate the attack patterns in the provenance graph because the text information of files and thread activities are too fine-granularityed for data provenance, which thus results in detection failure. It is a common limitation for provenance-based methods. Improving the granularity of data provenance or utilizing another novel data source to detect host-based threats are our future works.

%% file: conclusion.tex
\section{Conclusion}

\label{section:9}

We present \textsc{threaTrace}, a real-time host-based node level threats detection system that takes whole-system provenance graphs as input to detect and trace stealthy intrusion behavior. \textsc{threaTrace} uses a multi-model framework to learn different roles of benign nodes in a system based on their features and the causality relationships between their neighbor nodes, and detect stealthy intrusion behavior based on detecting abnormal nodes while locating them at the same time. The evaluation results show that \textsc{threaTrace} successfully detects and locates stealthy threats with high detection performance, processing speed, and acceptable system resource overhead. 